\documentclass[twocolumn,aps,superscriptaddress,showpacs,prb]{revtex4}
\usepackage{graphicx}
\usepackage{times}
\usepackage{amsmath}
\usepackage{amssymb}
\usepackage{units}
\usepackage{stmaryrd} 
\DeclareGraphicsExtensions{.pdf,.png, .eps ,.jpg}

\begin{document}
\preprint{0}

\title{The momentum and photon energy dependence of the circular dichroic photoemission in the bulk Rashba semiconductors BiTeX (X = I, Br, Cl)}

\author{A. Crepaldi}
\email{alberto.crepaldi@elettra.eu}
\affiliation{Elettra - Sincrotrone Trieste, Strada Statale 14 km 163.5, 34149 Basovizza-Trieste, Italy}

\author{F. Cilento}
\affiliation{Elettra - Sincrotrone Trieste, Strada Statale 14 km 163.5, 34149 Basovizza-Trieste, Italy}
\author{M. Zacchigna}
\affiliation{C.N.R. - I.O.M., Strada Statale 14 km 163.5 34149 Basovizza-Trieste, Italy}

\author{M. Zonno}
\affiliation{Universit\`a degli Studi di Trieste - Via A. Valerio 2 34127 Trieste, Italy}

\author{J. C. Johannsen}
\affiliation{Institute of Condensed Matter Physics (ICMP), Ecole Polytechnique
F\'ed\'erale de Lausanne (EPFL), CH-1015 Lausanne, Switzerland}

\author{C. Tournier-Colletta}
\affiliation{Institute of Condensed Matter Physics (ICMP), Ecole Polytechnique
F\'ed\'erale de Lausanne (EPFL), CH-1015 Lausanne,
Switzerland}

\author{L. Moreschini}
\affiliation{Advanced Light Source (ALS), Lawrence Berkeley National Laboratory, Berkeley, California 94720, USA}

\author{I. Vobornik}
\affiliation{C.N.R. - I.O.M., Strada Statale 14 km 163.5 34149 Basovizza-Trieste, Italy}

\author{F. Bondino}
\affiliation{C.N.R. - I.O.M., Strada Statale 14 km 163.5 34149 Basovizza-Trieste, Italy}

\author{E. Magnano}
\affiliation{C.N.R. - I.O.M., Strada Statale 14 km 163.5 34149 Basovizza-Trieste, Italy}

\author{H. Berger}
\affiliation{Institute of Condensed Matter Physics (ICMP), Ecole Polytechnique F\'ed\'erale de Lausanne (EPFL), CH-1015 Lausanne,
Switzerland}

\author{A. Magrez}
\affiliation{Institute of Condensed Matter Physics (ICMP), Ecole Polytechnique F\'ed\'erale de Lausanne (EPFL), CH-1015 Lausanne,
Switzerland}

\author{Ph. Bugnon}
\affiliation{Institute of Condensed Matter Physics (ICMP), Ecole Polytechnique F\'ed\'erale de Lausanne (EPFL), CH-1015 Lausanne,
Switzerland}

\author{G. Aut\`es}
\affiliation{Institute of Theoretical Physics, Ecole Polytechnique F\'ed\'erale de Lausanne (EPFL), CH-1015 Lausanne, Switzerland}

\author{O. V. Yazyev}
\affiliation{Institute of Theoretical Physics, Ecole Polytechnique F\'ed\'erale de Lausanne (EPFL), CH-1015 Lausanne, Switzerland}

\author{M. Grioni}
\affiliation{Institute of Condensed Matter Physics (ICMP), Ecole Polytechnique
F\'ed\'erale de Lausanne (EPFL), CH-1015 Lausanne,
Switzerland}

\author{F. Parmigiani}
\affiliation{Elettra - Sincrotrone Trieste, Strada Statale 14 km 163.5, 34149 Basovizza-Trieste, Italy}
\affiliation{Universit\`a degli Studi di Trieste - Via A. Valerio 2 34127 Trieste, Italy}

\date{\today}

\begin{abstract}

Bulk Rashba systems BiTeX (X = I, Br, Cl) are emerging as important candidates for developing spintronics devices, because of the coexistence of spin-split bulk and surface states, along with the ambipolar character of the surface charge carriers. The need of studying the spin texture of strongly spin-orbit coupled materials has recently promoted circular dichroic Angular Resolved Photoelectron Spectroscopy (cd-ARPES) as an indirect tool to measure the spin and the angular degrees of freedom.
Here we report a detailed photon energy dependent study of the cd-ARPES spectra in BiTeX (X = I, Br and Cl). Our work reveals a large variation of the magnitude and sign of the dichroism. Interestingly, we find that the dichroic signal modulates differently for the three compounds and for the different spin-split states. These findings show a momentum and photon energy dependence for the cd-ARPES signals in the bulk Rashba semiconductor BiTeX (X = I, Br, Cl). Finally, the outcome of our experiment indicates the important relation between the modulation of the dichroism and the phase differences between the wave-functions involved in the photoemission process. This phase difference can be due to initial or final state effects. In the former case the phase difference results in possible interference effects among the photo-electrons emitted from different atomic layers and characterized by entangled spin-orbital polarized bands. In the latter case the phase difference results from the relative phases of the expansion of the final state in different outgoing partial waves.

\end{abstract}

\maketitle


The need of novel and advanced spintronics devices has stimulated the quest for materials hosting metallic spin polarized bands embedded in a semiconducting bulk. Starting from the present knowledge on topological insulators (TIs) \cite{Fu_PRB_2007, FU_PRL_2007, Hsieh_Nature_2008, Hsieh_Science_2009, Hasan_RMP_2010}, the design of materials with spin-polarized bands requires the tailoring of the spin texture at the Fermi level ($E_{\rm F}$), hence the synthesis of systems such as the ternary TIs \cite{Chadov_NatMat_2010, Lin_NatMat_2010} or the bulk Rashba semiconductors BiTeX (X = I, Br and Cl) characterized by ambipolar surface states \cite{Ishi_NatMat_2011, Crepaldi_PRL_2012, Sakano_PRL_2013, Landolt_PRL_2012, Landolt_NJP_2013}. Nowadays one of the major challenge is to study the fully three-dimensional spin properties of ternary TIs, and the bulk Rashba semiconductors, as done for magnetic doped TIs \cite{Xu_NatPhys_2012}.

Spin resolved ARPES (sr-ARPES) offers the unique possibility to \emph{directly} address the spin polarization. Unfortunately, the sr-ARPES, based on high energy spin-dependent Mott scattering, is characterized by a low efficiency (10$^{-3}$--10$^{-4}$) \cite{Dil_rev_2009}. This limitation has recently renewed the interest for alternative spin detection devices based on higher efficiency low electron energy diffraction (IV-LEED with 10$^{-1}$--10$^{-2}$) \cite{okuda_Rev_2011}. This context well explains why the possibility of \emph{indirectly} studying the spin polarization via circular dichroic ARPES (cd-ARPES) was regarded as a major breakthrough \cite{Gedik_Rev_2013}. cd-ARPES measures the difference between the photoemission intensities obtained with the two opposite helicities of the circularly polarized light. However, there is no general consensus about the physical mechanism at the origin of the dichroism in cd-ARPES experiments. Recent theoretical and experimental studies propose to interpret the dichroism in giant spin split states in surface alloys \cite{Park_PRL_2011, Kim_PRB_2012, Guang_PRL_2012} and in TIs \cite{Wang_PRL_2011, Park_PRL_2012, Jung_PRB_2011, Mir_PRL_2012} as the result of local orbital angular momentum (OAM \cite{Park_PRL_2011, Bahramy_NC_2012, Park_PRB_2012, Park_PRL_2012, Kim_PRB_2012, Kim_arxiv_2013}) or directly as a measure of the spin polarization \cite{Wang_PRL_2011} . 

Recently, a large variation of the dichroism as a function of the incoming photon energy as well as the change in its sign have been reported by Scholz and coworkers for the $\mathrm{Bi_{2}Te_{3}}$ topological insulators \cite{Scholz_PRL_2013}. Similar observation has been reported in the TI $\mathrm{Bi_{2}Te_{2}Se}$ by Neupane et \emph{al.} \cite{Neupane_PRB_2013}. These works have shown that the dichroic signal cannot be interpreted only in terms of initial state effects, and that also final state effects must be accounted for \cite{Mulazzi_PRB_2009, Scholz_PRL_2013, Neupane_PRB_2013}. For addressing the origin of dichroism in photoemission, Zhu et \emph{al.} developed a model capable to explain the effect of linear dichroism in ARPES experiments on the TI $\mathrm{Bi_{2}Se_{3}}$ \cite{Zhu_PRL_2013}. They proposed that photoelectrons originating from different atomic layers in $\mathrm{Bi_{2}Se_{3}}$, and characterized by different spin and orbital projections of the wave-functions, can interfere.  
In this photoelectron interference process, the cd-ARPES signal modulates accordingly to a phase term which depends on several factors, namely the photon energy, the layer where the photoelectrons are originated (which translates into different optical paths) and on the particular orbital and spin projection of the electron wave-functions on those layers \cite{Zhu_PRL_2013}.
Recently, \"Arr\"al\"a and coworkers proposed a different description for the photon energy dependence of the dichroism. On the basis of relativistic photoemission calculations for the Au(111) surface states, they ascribed this modulation to a phase difference between the complex expansion of the final state wave-function over various partial waves, characterized by different orbital quantum numbers \cite{Lindroos_PRB_2013}. Interestingly, both the model of Zhu and of \"Arr\"al\"a predict that the dichroic signal is strongly modulated as a function of a phase difference, which is attributed respectively to a property of the initial or of the final states.


\begin{figure}[ttt]
 \includegraphics[width = 0.3 \textwidth]{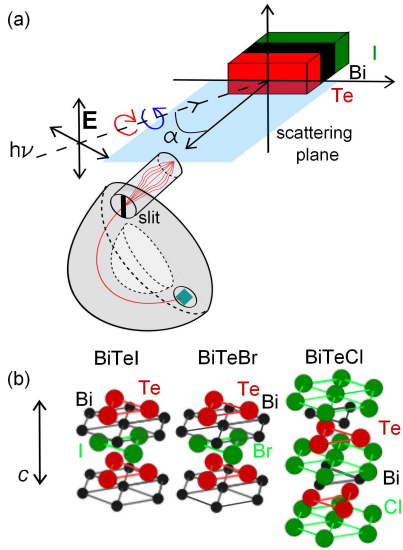}
  \caption[didascalia]{(Color online) (a) Crystal structure of BiTeX (X = I, Br and Cl) obtained from x-ray powder diffraction \cite{sheve_XRAY_1995}. (b) experimental geometry common to both ARPES end-stations. The analyzer slit is orthogonal to the scattering plane, defined by the photon and photo-electron wave-vectors.
   }
  \label{fig:arpes1}
\end{figure}

Here we report a detailed experimental investigation of the cd-ARPES signal measured in bulk Rashba compounds, BiTeX (X = I, Br and Cl). The experimental data show a large variation of the dichroic signal as a function of the impinging photon energy. Furthermore, the photon energy dependence of the cd-ARPES signal varies in the same spin polarized band for different regions of its parabolic dispersion, unveiling a momentum dependence of the photon-energy modulation of the dichroism. The BiTeBr and the BiTeCl compounds, in contrast to BiTeI, host several spin split states crossing $E_{\rm F}$. Recently, this behavior has been ascribed to states located at different depths below the surface and resulting from a staircase-like potential \cite{Landolt_NJP_2013}. Interestingly, we observe that the dichroic signal modulates with the photon energy differently in these various states. We propose that the momentum and photon energy dependence of the dichroism reflects a phase difference between the outgoing partial waves, as recently proposed by \"Arr\"al\"a et \emph{al.} for the Au(111) surface state \cite{Lindroos_PRB_2013} or, alternatively, that it results from photoelectron interference effects, as proposed by Zhu and co-workers in the case of the TI $\mathrm{Bi_{2}Se_{3}}$ \cite{Zhu_PRL_2013}.


cd-ARPES experiments have been carried out on the BiTeX (X = I, Br, Cl) compounds at the APE beamline at the Elettra synchrotron in the energy range between 20~eV and 50~eV. The UV light was generated by an APPLE II undulator with a high degree of circular polarization  at the sample position ($>$~90~$\%$ \cite{Pana_RSI_2009}). The ARPES end-station was equipped with a Scienta SES 2002 analyzer, with an overall energy and angular resolution respectively set to 15~meV and 0.2$^{\circ}$. The samples were cleaved \emph{in situ} at room temperature and measured at liquid nitrogen temperature ($\sim$~77~K). A set of measurements at higher photon energy, in the range 80--180 eV, was performed on BiTeI at the BACH beamline, at the Elettra synchrotron. The measurements were performed at liquid nitrogen temperature with the use of a Scienta R3000 analyzer with an overall energy and angular resolution set to 27~meV and 0.1$^{\circ}$. Also this beamline provides us with a high degree of circular polarization ($>$~99.7~$\%$) through the use of an APPLE II undulator \cite{Zangrando_RSI_200431}.

High quality BiTeBr crystals were grown by chemical vapor transport. Stoichiometric mixture of Bi, Te and $\mathrm{BiBr_{3}}$ were sealed with HBr as transport agent. The ampule was placed in a two-zone furnace with charge and growth temperature 440$^{\circ}$~C and 400$^{\circ}$~C, respectively.
BiTeI crystals were produced by melting in a sealed quartz ampule a stoichiometric mixture of Bi, Te and $\mathrm{BiI_{3}}$ at 600$^{\circ}$C. The horizontal furnace is subsequently cooled to 200$^{\circ}$C at a rate of 1K/h.  The synthesis of BiTeCl was realized from $\mathrm{Bi_{2}Te_{3}}$ with $\mathrm{BiCl_{3}}$ in excess ($\mathrm{BiCl_{3}}$/$\mathrm{Bi_{2}Te_{3}}$ $>$ 5). The quartz ampule was placed vertically inside a muffle furnace. During the growth, the temperature at the bottom and at the top of the ampule were respectively 440$^{\circ}$~C and 400$^{\circ}$~C. The temperatures were maintained for few days. Then the furnace was cooled down to room temperature at 1K/h. At the end of the processes, centimeter large crystals were obtained. The structure and chemical composition were confirmed by x-ray diffraction and energy dispersive x-ray spectroscopy.


\begin{figure}[ttt!]
 \includegraphics[width = 0.5 \textwidth]{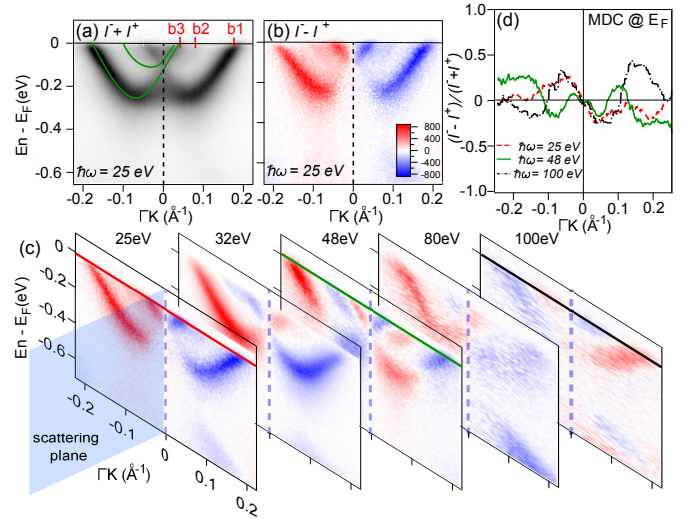}
  \caption[didascalia]{(Color online) (a) Band dispersion of BiTeI measured at 25~eV along the $\overline{\Gamma K}$ high symmetry direction and (b) corresponding dichroic signal ($I^{-} - I^{+}$). Dashed vertical line indicates the scattering plane at $k~=~0~\AA^{-1}$, green lines are eye-guide for the parabolic dispersion of the left spin branches. (c) Evolution of the dichroism in BiTeI at selected photon energies (25~eV, 32~eV, 48~eV, 80~eV, 100~eV). (d) MDCs at $E_{\rm F}$ for 25~eV, 48~eV and 100~eV respectively in red, green and black. The number of sign changes (zero crossing) is respectively equal to one, five and three.
   }
  \label{fig:arpes1}
\end{figure}


Figure~1~(a) shows the geometry common to both ARPES end-stations. The incoming photon wave-vector and the out-going photoelectron wave-vector form an angle $\alpha~=~45^{\circ}$ ($\alpha~=~60^{\circ}$ on the beamline BACH) and they define the scattering plane. The circular polarization of the light is transform from left to right by a mirror like symmetry about the scattering plane. In the case of photoelectron wave-vectors lying in the scattering plane the dichroic signal must be zero, unless the compound breaks time-reversal symmetry \cite{Borisenko_PRL_2004, Scholz_PRL_2013, Ishida_PRL_2011}, which is not the case for the materials here investigated. In order to resolve a dichroic signal the setup must possess a \emph{handedness}, \emph{i.e.} the photoelectron wave-vectors should not lie in the mirror plane. In both the experimental end-stations the analyzer slit is orthogonal to the experimental scattering plane, thus allowing for a direct measurement of the circular dichroism. It is well known that the experimental geometry can also influence the circular dichroism \cite{Ishida_PRL_2011, Gedik_Rev_2013}, and a dichroic signal can be introduced also by the experimental \emph{handedness}. In order to disentangle this \emph{artificial} effect from the physical properties of the material, several in-plane orientations of the BiTeI were measured, and the dichroic signal rotates accordingly (not shown). This proves that the dichroism does not originate from asymmetry in the experimental setup, but it arises from physical properties inherent to the materials.


Figure~1~(b) shows the crystal structure of the three BiTeX systems as determined by x-ray diffraction \cite{sheve_XRAY_1995}. All the compounds are non-centrosymmetric small gap semiconductors described within the semi-ionic model \cite{sheve_XRAY_1995}. In the case of BiTeI, the unit cell is formed by the alternation of Bi, Te and I layers, with $P3m1$ point group symmetry. The natural cleavage plane is between Te and I and both terminations are possible \cite{Crepaldi_PRL_2012, Cedric_PRB_2014}. In this study we focus only on the electronic properties of the Te-terminated surfaces of all the compounds. Our x-ray diffraction study indicates that in BiTeBr the Te and Br atoms are well organized in separated and alternating layers, similarly to BiTeI, and they do not form a mixed alloy as originally proposed \cite{sheve_XRAY_1995}. A similar structural model has been recently proposed by ARPES experiments \cite{Sakano_PRL_2013} supported by theoretical calculations \cite{Rusinov_PRB_2013}. The crystal structure of BiTeCl differs as it is characterized by quintuple layers of alternating Bi, Te and Cl layers with $P6_3mc$ point group symmetry, and the unit cell parameter $c$ is doubled ($c~=~12.39~\AA$) with respect to the one of BiTeI ($c~=~6.85~\AA$) and BiTeBr ($c~=~6.48~\AA$) \cite{sheve_XRAY_1995}.


Figure~2~(a) displays the electronic band structure of BiTeI measured with photon energy equal to 25~eV along the $\overline{\Gamma K}$ high symmetry direction ($\overline{\Gamma K} = 0.96~\AA^{-1}$). The figure results from the sum of the data measured with the two circular polarizations. Several spectral features are observed, in excellent agreement with previous ARPES studies \cite{Ishi_NatMat_2011, Crepaldi_PRL_2012, Landolt_PRL_2012, Sakano_RPB_2012}. Dashed vertical line indicates the scattering plane at $k~=~0~\AA^{-1}$, green lines are eye-guide for the parabolic dispersion of the left spin branches. The bands crossing at the Fermi level for positive \emph{k} is indicated by red marks. We associated the \emph{b1} and \emph{b3} states to the outer and inner branches of the spin polarized surface states. An additional state, \emph{b2},  disperses between the two. It is hardly detectable at this photon energy and we attribute it to the outer branch of the three-dimensional bulk derived spin split state \cite{Crepaldi_PRL_2012, Landolt_PRL_2012}. Fig.~2~(b) shows the cd-ARPES image ($I^{-} - I^{+}$) associated to Fig.~2~(a). At 25~eV photon energy the dichroism changes sign along the parabolic dispersion of each spin branch at the scattering plane. The dichroic signal is always positive (negative) for positive (negative) \emph{k}, similarly to what reported very recently by a low photon energy cd-ARPES study of BiTeI \cite{Perfetti_PRL_2013}. Surprisingly, this is not the case for all the photon energies, as clearly shown in Fig.~2~(c). A small variation of the photon energy (from 25~eV to 32~eV) is sufficient to induce a dramatic change in the dichroism of the \emph{b3} state (the inner contour) which has now opposite sign. Hence, at 32~eV photon energy the dichroic signal does not change sign along the parabolic dispersion of the surface state. Therefore, at this particular energy the dichroic signal of the spin split surface states mimics the expected spin polarization of the initial state. Conversely, the complete change in the dichroism for the \emph{b1} states (the outer branch) is achieved only at $\sim$~100~eV. The k-separation between the \emph{b1} and \emph{b3} states is small (it corresponds to their spin splitting, $2k_0 \sim 0.11~\AA^{-1}$) and the weak dispersion of the final states is not likely to account for such a large difference in the modulation of the dichroism as a function of the photon energy. 



\begin{figure}[t!]
 \includegraphics[width = 0.5 \textwidth]{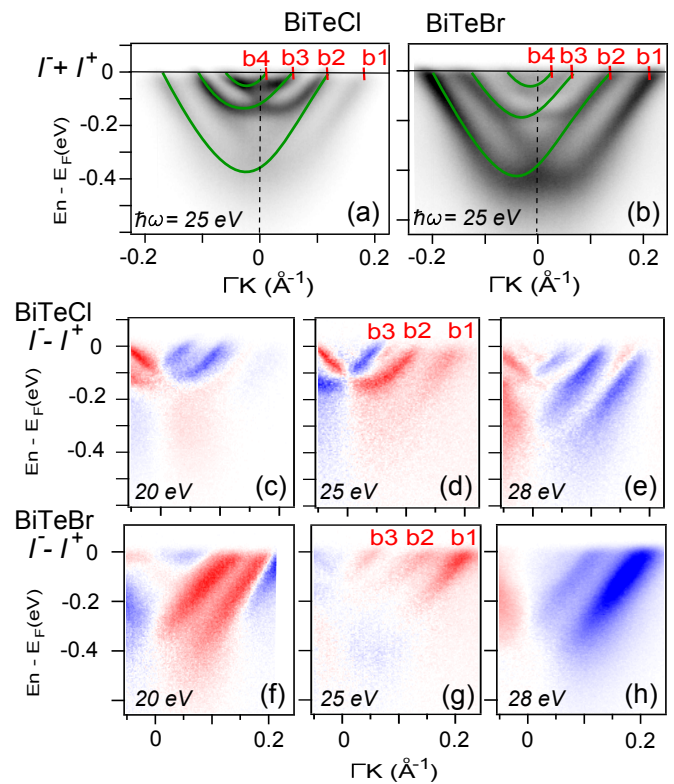}
  \caption[didascalia]{(Color online) (a), (b) Electronic band dispersion of BiTeCl and BiTeBr, respectively,  measured along the $\overline{\Gamma K}$ high symmetry direction with photon energy equal to 25~eV. Dashed vertical line indicates the scattering plane at $k~=~0~\AA^{-1}$, green lines are eye-guide for the parabolic dispersion of the left spin branches. (c) - (h) dichroic maps at 22~eV, 25~eV and 28~eV for the two materials. The dichroic signal in the states labelled \emph{b1} - \emph{b4}, whose $k_{\rm F}$ is marked by red tick marks, is different in the two compounds. 
  }
  \label{fig:arpes2}
\end{figure}


At 48~eV photon energy the ARPES signal differs significantly from the one recorded at lower photon energies, and the bulk state \emph{b2} is more clearly resolved.  Interestingly, the dichroism changes sign along the band at $\sim ~ 0.1~\AA ^{-1}$, which corresponds approximately to Fermi wave-vector of \emph{b2}. The abrupt change in sign of the dichroism is suggestive of a hybridization between the surface derived states and the bulk state. This hybridization is observed at 48~eV owing to the three dimensional character of the bulk state and it can be responsible for a partial re-orientation of the spin and orbital projections. In BiTeCl, for example, the theory shows that the spin polarization along the \emph{z} direction changes when the surface state approaches the bulk conduction band \cite{Eremev_PRL_2012}.

The different behavior of the dichroism at the three energies is summarized in Fig.~2~(d), where the momentum distribution curves (MDCs) at the Fermi level ($E_{\rm F}$) are extracted from the cd-ARPES images of Fig.~2~(c). At 25~eV photon energy (red) the dichroism changes sign only at the scattering plane ($k = 0~\AA ^{-1}$). Instead, at 100~eV (black) and 48~eV (green) photon energies we observe three and five sign changes, respectively. This shows that the dichroism modulates with the photon energy differently for different value of the electron momentum.

The momentum and photon energy dependence of the cd- ARPES signal is investigated also in BiTeBr and BiTeCl. Recent high resolution ARPES studies \cite{Sakano_PRL_2013, Landolt_NJP_2013} and theoretical calculations \cite{Rusinov_PRB_2013} reported the existence of several surface derived spin-split states at the Te termination of BiTeBr and BiTeCl, in contrast to the unique pair of surface states observed in BiTeI. Density functional theory (DFT) calculations show that each spin-split state lies at different depth below the surface and it is confined between different crystal unit cells \cite{Landolt_NJP_2013}. Therefore, it is interesting to verify whether dichroism modulates differently in the various bands of these two compounds.

Figure~3~(a) and (b) display the high resolution ARPES data of the electronic band structure of BiTeCl and BiTeBr measured along the $\overline{\Gamma K}$ high symmetry direction at 25~eV photon energy. In BiTeCl and BiTeBr the spin branches disperse with parabolic behavior and similar positive effective mass in an extremely small region around the $\Gamma$ point of the Brillouin Zone (BZ). The outer spin branch of each parabola is found to be almost degenerate with the inner spin branch of the successive band \cite{Sakano_PRL_2013}. All those spin-split states are ascribed to the polar nature of the semi-ionic crystals and to the resulting staircase-like potential landscape \cite{Landolt_NJP_2013}. 

For BiTeCl we did not observe the linearly dispersing topologically protected surface state recently reported by Chen et \emph{al.} \cite{Chen_NPh_2013}. Instead, our data indicate the existence of several spectral features associated to spin-split states localized in different unit cells beneath the surface in good agreement with other independent studies \cite{Sakano_PRL_2013, Landolt_NJP_2013}. Green guide lines trace the parabolic dispersion of the left spin branches. We indicate the Fermi wave-vectors for positive \emph{k} with red tick marks and labels \emph{b1} - \emph{b4}. Unfortunately, we cannot resolve the contributions of the inner and outer spin branches of the two consecutive states, because of their small \emph{k}-separation. Nevertheless, the evolution of the dichroic signal \emph{averaged over diverse states} is sufficient to capture the different photon-energy modulation of the dichroism in the different region of the Brillouin zone.

In Fig.~3~(c)-(h) we report the cd-ARPES images for BiTeCl (central panels) and for BiTeBr (bottom panels) respectively at 22~eV, 25~eV and 28~eV photon energies. In both materials the dichroic signal associated to the most external branch, \emph{b1}, displays a smaller modulation with the photon energy. This is in agreement with what was found in BiTeI, where the change in sign is observed only at $\sim$ 100~eV photon energy. The states with smaller $k_{\rm F}$ are instead characterized by a more frequent modulation. In particular in BiTeCl the \emph{b2} band switches sign twice as the photon energy is varied over a window of $\sim$ 8 eV. Besides the evolution of the dichroism in different states in the same compound, it is interesting to note the difference of the dichroic signal in the same state between the two compounds. In particular, the sign of the dichroism in the two systems is opposite in \emph{b3} at 25~eV, whereas it is opposite in \emph{b1} and \emph{b2} at 22 eV and it is the same in all the states when measured at 28~eV.



\begin{figure}[t!]
 \includegraphics[width = 0.45\textwidth]{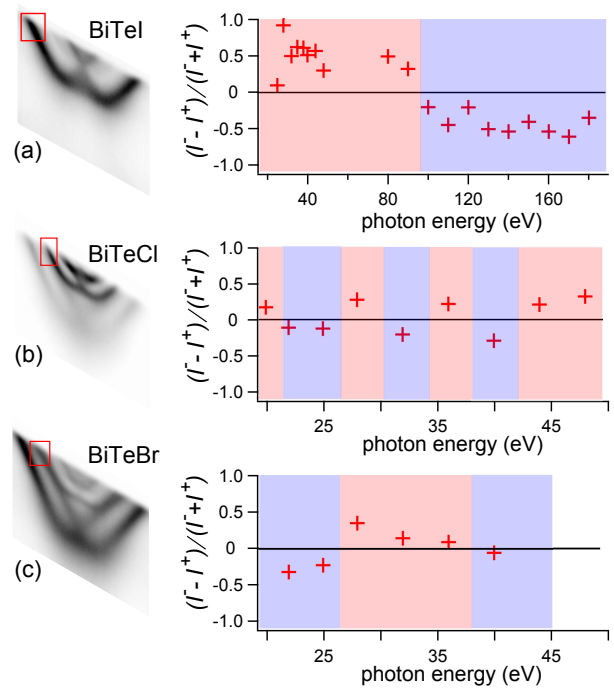}
  \caption[didascalia]{ (Color online) evolution of the normalized dichroic signal $(I^{-} - I^{+}) / (I^{-} + I^{+})$ for (a) BiTeI (b) BiTeCl  and (c) BiTeBr. Markers report the averaged signal at the Fermi level for the bands indicated by red rectangles.  
 }
  \label{fig:arpes3}
\end{figure}


In order to compare more quantitatively the modulation of the dichroic signal in the different compounds we report in Figure~4 the normalized dichroism, defined as $(I^{-} - I^{+}) / (I^{-} + I^{+})$, for BiTeI (panel~a) BiTeCl (panel~b) and BiTeBr (panel~c). Each panel shows red markers reporting the averaged signal at the Fermi level integrated over the momentum region enclosed by the rectangle. The color of the background, instead, highlights the change in the sign of the dichroism. 

We reveal only one change in sign in BiTeI at $\sim$ 100~eV, whereas in the case of BiTeCl the dichroism changes sign six times in an energy window $<$ 30~eV. The trend of BiTeBr is somewhere in between, and we report two changes of sign in the measured energy range (25~eV). Furthermore, BiTeI is characterized by the largest dichroic signal, which is on average larger than $50~\%$ and it reaches almost $92~\%$ at 28~eV, while for the two other compounds the dichroic signal is always found smaller than $25~\%$. In general, we observe a smaller number of sign changes in the dichroic signal associated to the states having larger Fermi wave-vectors, one (two) sign change less for BiTeBr (BiTeCl). 


Variation of the cd-ARPES signal as a function of the photon energy has been already reported for the TI $\mathrm{Bi_{2}Te_{3}}$ and this effect has been interpreted in term of final state effects \cite{Scholz_PRL_2013}. In particular calculations of the ARPES intensity in one-step model suggest that the magnitude of the circular dichroism associated to the \emph{p}-like surface state is strongly affected when the photon energy matches a final state with a large \emph{d}-like contribution, due to the selection rules for the orbital angular momentum in an optical electric dipole transition. Nevertheless, in the present study, in stark contrast to the single spin polarized Dirac particle at the surface of $\mathrm{Bi_{2}Te_{3}}$ \cite{Scholz_PRL_2013}, the bulk Rashba semiconductors BiTeX display several bands close in energy and momentum. Owing to the small momentum and energy distance between these states, and owing to the weak dispersion of the final state, the dipole selection rule alone can hardly account for the observed momentum dependence of the circular dichroism.

We propose that the observed momentum and photon energy dependence of the cd-ARPES signal in BiTeX could be a consequence of the difference in the phase terms of the wave-functions, describing the initial \cite{Zhu_PRL_2013} and/or final states \cite{Lindroos_PRB_2013}. This phase term could be responsible for photoelectron interference effects, as proposed by Zhu and coworkers \cite{Zhu_PRL_2013}. In the model of Zhu this phase difference results from the photon energy, the layer where the photoelectrons are originated (which translates into different optical paths) and the orbital and spin layer-projection of the wave-functions \cite{Zhu_PRL_2013}.
For the \"Arr\"al\"a model the phase change results from the phase difference between the complex expansion of the final state on outgoing partial waves characterized by different orbital quantum number. Detailed calculations accounting for  both initial and final state effects and the relative phase difference between the photo-electron wave-functions are required to clarify the role of the different terms.


In summary we have performed a photon energy dependent study of the circular dichroic ARPES signal in the bulk Rashba materials BiTeX (X = I, Br and Cl). We report on the modulation of the dichroism as a function of the incoming photon energy. We have observed that the evolution of the dichroic signal in BiTeI varies along the parabolic dispersion of each spin branch. In the case of BiTeCl and BiTeBr several surface derived spin split states are observed, in agreement with the literature. Our data show that also in these compounds the circular dichroism modulates differently as a function of the photon energy in the various states, thus indicating a momentum and photon energy dependence of the circular dichroism.

We gratefully acknowledge A. Damascelli and Z.-H. Zhu for discussions. I.V., F.B. and E.M. acknowledge the technical support by F. Salvador and P. Bertoch (CNR-IOM). This work was supported in part by the Italian Ministry of University and Research under Grant Nos. FIRBRBAP045JF2 and FIRB-RBAP06AWK3 and by the European Community–Research Infrastructure Action under the FP6 "Structuring the European Research Area" Programme through the Integrated Infrastructure Initiative "Integrating Activity on Synchrotron and Free Electron Laser Science" Contract No. RII3-CT-2004-506008.  G.A. and O.V.Y. were supported by the Swiss NSF grant No.~PP00P2\_133552 and the ERC starting grant ``TopoMat'' (No.~306504). F.B. and E.M. acknowledge the support by the Italian MIUR in part through the national grant Futuro in ricerca 2012 RBFR128BEC.


\end{document}